\documentclass[preprint,aps]{revtex4}

\usepackage{graphics}

\begin{document}

\title{The Onset of Chaos in Spinning Particle Models}

\author{H. T. Cho}
  \email{htcho@mail.tku.edu.tw}
\author{J.-K. Kao}
  \email{g3180011@tkgis.tku.edu.tw}
\affiliation{Department of Physics, Tamkang University, Tamsui,
Taipei, Taiwan, Republic of China}

\date{\today}

\begin{abstract}

The onset of chaos in one-dimensional spinning particle models
derived from pseudoclassical mechanical hamiltonians with a
bosonic Duffing potential is examined. Using the Melnikov method,
we indicate the presence of homoclinic entanglements in models
with general potentials for the spins, and thus show that chaotic
motions occur in these models.

\end{abstract}


\maketitle

Suzuki and Maeda \cite{suzuki} has recently showed that the
motions of spinning particles in the Schwarzschild black hole
spacetime could be chaotic. This is an important result due to its
relevancy to the detection of gravitational waves from black hole
coalescences \cite{levin,cornish}, and it has also aroused our
interest in considering the chaotic motions of spinning particles.

On the other hand, spinning particle motions have been considered
in \cite{holten1,holten2} from a different point of view. The
authors there described spinning particles as particles moving in
a supersymmetric space, which they called the spinning space, with
Grassmann-valued vector variables in addition to the ordinary
spacetime coordinates. Hence, the interplay between chaos and
spinning particle models involving pseudoclassical mechanics
\cite{casalbuoni} looks rather intriguing.

In this letter we would like to consider the simplest
one-dimensional spinning particle model (Eq.~(\ref{g3model})
below) derived from pseudoclassical mechanical systems . This is
just the spinning particle model considered by Berezin and Marinov
\cite{berezin}, or the so-called $G_{3}$ model according to the
classification scheme in \cite{casalbuoni}. We hope that the
investigation of this simple model will shed some light on the
chaotic behaviors of more complicated spinning particle models in
higher dimensions or even in curved spaces.

Here we use an analytic method called the Melnikov technique
\cite{melnikov,holmes,bombelli} to detect the onset of chaotic
motions. Suppose in an integrable system there are homoclinic
orbits emanating and terminating at the same unstable fixed point.
Time-dependent perturbations are then added to this system. In the
perturbed system the stable and unstable orbits split. An integral
called the Melnikov function evaluated along the unperturbed
homoclinic orbit measures the transversal distance between the
perturbed stable and unstable orbits on the Poincar\'e section.
The presence of isolated zeros in the Melnikov function indicates
complicated entanglements between the two perturbed orbits and
thus the presence of chaotic behaviors.

We start with the bosonic part of our model. We choose the
one-dimensional Duffing hamiltonian,
\begin{equation}
H=\frac{p^2}{2}+V(x),
\end{equation}
where
\begin{equation}
V(x)=-x^2+x^4.
\end{equation}
The equations for the homoclinic orbit (separatrix) with energy
$E=0$ are particularly simple,
\begin{eqnarray}
x_{s}(t)&=&{\rm sech}({\sqrt{2}}t),\\ p_{s}(t)&=&-\sqrt{2}\ {\rm
sech}({\sqrt{2}}t)\ {\rm tanh}({\sqrt{2}}t).
\end{eqnarray}
This is the main reason we choose the Duffing potential. On the
other hand, we expect similar analysis as the one we carry out
below can be applied to other potentials with homoclinic or
heteroclinic orbits.

To consider pseudoclassical models with Grassmann variables, we
concentrate on those with three Grassmann variables $\xi_{1}$,
$\xi_{2}$, and $\xi_{3}$,
\begin{equation}
H=\frac{p^2}{2}+V(x)-
\frac{i}{2}\sum^{3}_{i,j,k=1}\epsilon_{ijk}W_{i}(x)\xi_{j}\xi_{k}.
\label{g3model}
\end{equation}
where $W_{i}(x)$ are the potentials for the Grassmann variables.
The hamiltonian can be written as \cite{casalbuoni}
\begin{equation}
H=\frac{p^2}{2}+V(x)-\vec{W}(x)\cdot\vec{S}\label{g3}
\end{equation}
where the ``spin" $\vec{S}$ is defined by
\begin{equation}
S_{i}=-\frac{i}{2}\epsilon_{ijk}\xi_{j}\xi_{k}. \label{defspin}
\end{equation}
The dynamical equations are
\begin{eqnarray}
\frac{dx}{dt}&=&p,\\ \frac{dp}{dt}&=&-V'(x)+\vec{W}'(x)\cdot
\vec{S}, \label{tdpert}\\
\frac{d\vec{S}}{dt}&=&-\vec{W}(x)\times\vec{S}, \label{spineq}
\end{eqnarray}
where the prime $'$ denotes $d/dx$. Note that from here on, we
work directly with this set of dynamical equations without further
reference to the Grassmann nature of $\vec{S}$ as defined in
Eq.~(\ref{defspin}). That is, we {\em do not} restrict
\begin{equation}
S_{i}^{2}\sim 0.
\end{equation}
Hence, we are dealing with the spinning particle models defined by
the dynamical equations rather than the general pseudoclassical
systems with Grassmann variables \cite{Manton}. In effect we just
use the pseudoclassical hamiltonians as a convenient and
systematic means to derive the dynamical equations for the
spinning particles.

To detect the onset of chaotic motions in the spinning particle
models above, we treat the spin $\vec{S}$ as a perturbation. That
is, we assume that
\begin{equation}
x,p\sim O(1)\ \ {\rm and}\ S_{i}\sim O(\epsilon)
\end{equation}
for some small parameter $\epsilon$. Then the additional spin
terms in Eq.~(\ref{tdpert}) have the form of time-dependent
perturbations which may trigger chaotic behaviors.

\begin{figure}[!]
\resizebox*{3.0in}{2.0in}{\includegraphics{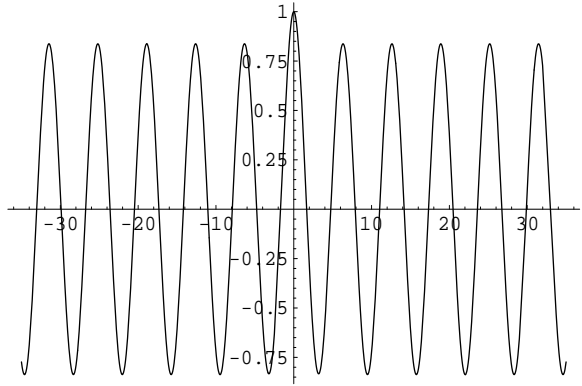}}
\caption{\label{spin1} Time evolution of $\tilde S_{1}$ in
Eqs.~(\ref{tils1})-(\ref{tils3}) with the initial conditions as
given in Eq.~(\ref{initial}).}
\end{figure}
\begin{figure}[!]
\resizebox*{3.0in}{2.0in}{\includegraphics{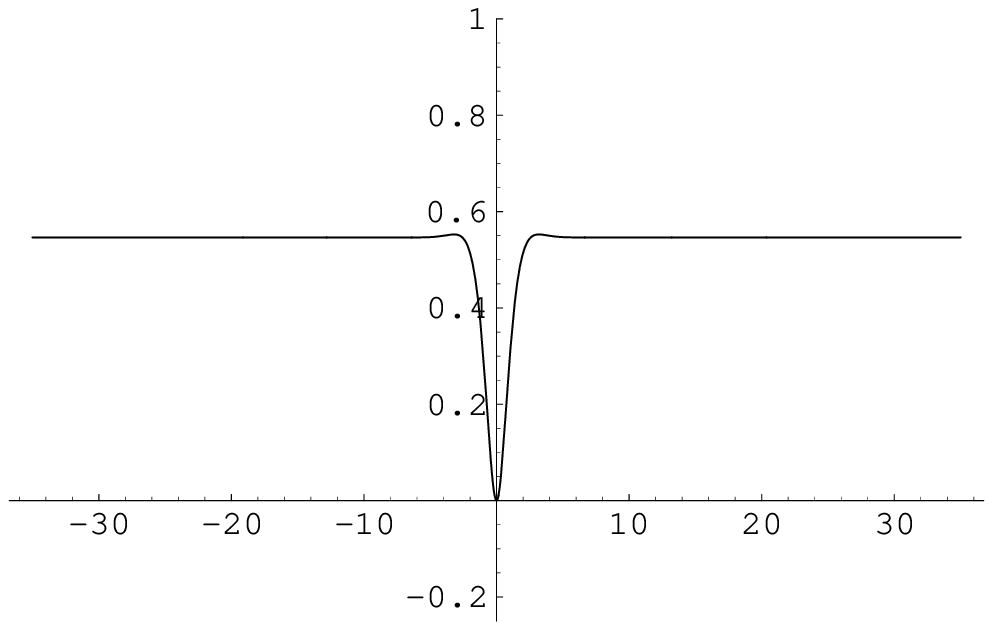}}
\caption{\label{spin2} Time evolution of $\tilde S_{2}$ in
Eqs.~(\ref{tils1})-(\ref{tils3}) with the initial conditions as
given in Eq.~(\ref{initial}).}
\end{figure}
\begin{figure}[!]
\resizebox*{3.0in}{2.0in}{\includegraphics{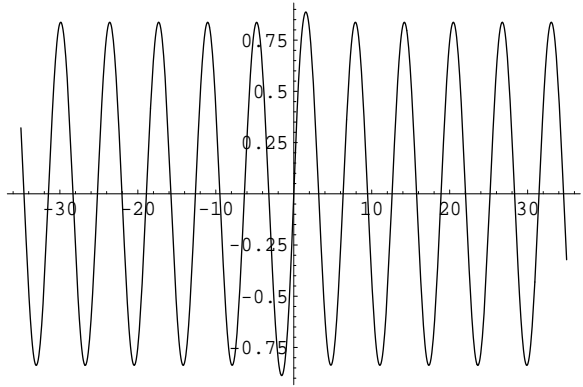}}
\caption{\label{spin3} Time evolution of $\tilde S_{3}$ in
Eqs.~(\ref{tils1})-(\ref{tils3}) with the initial conditions as
given in Eq.~(\ref{initial}).}
\end{figure}

To make the discussion more concrete we first take a simple form
for the potentials $\vec{W}(x)$,
\begin{equation}
W_{1}=x\ ;\ W_{2}=1\ ;\ W_{3}=0.\label{simple}
\end{equation}
We shall return to the consideration of more general $\vec{W}(x)$
below. To obtain the lowest order solutions of $S_{i}$, we just
replace $\vec{W}(x)$ in the spin dynamical equations
(Eq.~(\ref{spineq})) by $\vec{W}(x_{s}(t))$, that is,
\begin{eqnarray}
\frac{d{\tilde S}_{1}}{dt}&=&-{\tilde S}_{3}\label{tils1}\\
\frac{d{\tilde S}_{2}}{dt}&=&{\rm sech}({\sqrt{2}}t){\tilde
S}_{3}\label{tils2}\\ \frac{d{\tilde S}_{3}}{dt}&=&{\tilde
S}_{1}-{\rm sech}({\sqrt{2}}t){\tilde S}_{2}\label{tils3}
\end{eqnarray}
where we have written $S_{i}=\epsilon \tilde S_{i}$. We can
numerically solve this set of differential equations quite easily
using, for example, Mathematica. The solutions with the initial
conditions
\begin{equation}
{\tilde S_{1}}(0)=1\ \ \ ;\ \ \ {\tilde S_{2}}(0)={\tilde
S_{3}}(0)=0,\label{initial}
\end{equation}
are plotted in Figs.~\ref{spin1}-\ref{spin3}. $\tilde S_{1}$ and
$\tilde S_{3}$ are oscillatory, while $\tilde S_{2}$ is mostly
constant except for a small region near $t=0$. Note that the
choice of the initial conditions here is quite arbitrary. We have
checked the behaviors of the spins for other sets of initial
conditions like ${\tilde S_{1}}(0)=0$; ${\tilde S_{2}}(0)=1$;
${\tilde S_{3}}(0)=0$ and ${\tilde S_{1}}(0)=0$; ${\tilde
S_{2}}(0)=0$; ${\tilde S_{3}}(0)=1$, and similar results like
those in Figs.~\ref{spin1}-\ref{spin3} are obtained.

\begin{figure}[t]
\resizebox*{3.0in}{2.0in}{\includegraphics{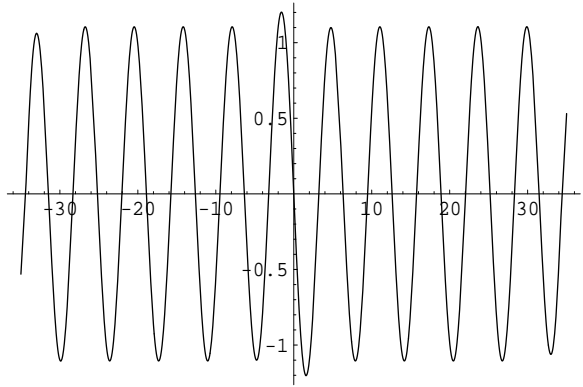}}
\caption{\label{melnikov} Melnikov function of the spinning
particle model in Eq.~(\ref{g3}) with the potentials for the spins
$\vec{W}(x)$ as given in Eq.~(\ref{simple}) .}
\end{figure}

It is well known that external oscillatory perturbations may
induce chaotic behaviors. We can see that this is indeed the case
here by calculating the Melnikov function \cite{melnikov,holmes},
\begin{eqnarray}
M(t_{0})&=&\int^{\infty}_{-\infty}dt\
p_{s}(t+t_{0})\vec{W}'(x_{s}(t))\cdot\vec{S}(t)\nonumber\\
&=&-\int^{\infty}_{-\infty}dt\ {\sqrt{2}}\ {\rm
sech}[{\sqrt{2}}(t+t_{0})]\ {\rm
tanh}[{\sqrt{2}}(t+t_{0})]S_{1}(t),\label{melnikovdef}
\end{eqnarray}
where $t_{0}$ is used to parametrize the location along the
unperturbed homoclinic orbit. $M(t_{0})$ is plotted in
Fig.~\ref{melnikov}. The Melnikov function measures the transveral
distance on the Poincar\'e section between the perturbed stable
and unstable orbits passing through the unstable fixed point. The
infinite number of zeros as seen in Fig.~\ref{melnikov} indicates
the entanglement of the two orbits, and this signals the
occurrence of chaotic behavior for the perturbed homoclinic
orbits. When the perturbation terms, or the spin terms here,
become larger and larger, the chaotic behavior will spread to
other parts of the phase space.

For more general $\vec{W}(x)$ potentials, we can argue that
similar chaotic behaviors as those with the simple potential in
Eq.~(\ref{simple}) discussed above will again occur. In this case
the Melnikov function is still given by Eq.~(\ref{melnikovdef})
but with general $\vec{W}(x)$. The key point to the argument is to
consider the large $t_{0}$ behavior of $M(t_{0})$. This is just
the region near the unstable fixed point where complicated
homoclinic entanglements may occur. Now in the integrand in
Eq.~(\ref{melnikovdef}),
\begin{equation}
p_{s}(t+t_{0})=-{\sqrt{2}}\ {\rm sech}[{\sqrt{2}}(t+t_{0})]\ {\rm
tanh}[{\sqrt{2}}(t+t_{0})].
\end{equation}
which is exponentially small except for
\begin{equation}
t\sim -t_{0}.
\end{equation}
Hence, to estimate $M(t_{0})$ for large $t_{0}$, we need only to
consider the behaviors of $\vec{W}'(x_{s}(t))$ and $\vec{S}(t)$
for $t\sim -t_{0}$ or for large $|t|$.

For large $|t|$, $x_{s}(t)$ is also exponentially small. Suppose,
for small $x$,
\begin{equation}
W_{i}\simeq a_{i}+b_{i}x+\cdots, \label{smallx}
\end{equation}
where $a_{i}$ and $b_{i}$ are constants. Then, for large $|t|$,
$W_{i}$ can be approximated by $a_{i}$, and the spin equations
become
\begin{equation}
\frac{d}{dt}\left(
\begin{array}{c}
{\tilde S}_{1}\\ {\tilde S}_{2}\\ {\tilde S}_{3}
\end{array}\right)
\simeq \left(
\begin{array}{ccc}
0&a_{3}&-a_{2}\\ -a_{3}&0&a_{1}\\ a_{2}&-a_{1}&0
\end{array}\right)
\left(
\begin{array}{c}
{\tilde S}_{1}\\ {\tilde S}_{2}\\ {\tilde S}_{3}
\end{array}\right).
\end{equation}
The eigenvalues of the matrix above are $0$ and $\pm ia$, where
$a= \sqrt{a_{1}^{2}+a_{2}^{2}+a_{3}^{2}}$. The solutions of
${\tilde S}_{i}$ are therefore linear combinations of ${\rm
sin}(at)$, ${\rm cos}(at)$, and some constants. Except for the
case when all $a_{i}=0$, some of the components ${\tilde S}_{i}$
must be oscillatory.

Similarly, for large $|t|$,
\begin{equation}
{W'}_{i}\simeq b_{i}.
\end{equation}
Hence the Melnikov function for large $t_{0}$ can be written
approximately as a linear combination of
\begin{widetext}
\begin{eqnarray}
M_{1}(t_{0})&=&\int^{\infty}_{-\infty}dt\ p_{s}(t+t_{0})=0,\\
M_{2}(t_{0})&=&\int^{\infty}_{-\infty}dt\ p_{s}(t+t_{0})\ {\rm
sin}(at)=-\frac{\pi a}{\sqrt{2}}\ {\rm sech}\!\left(\frac{\pi
a}{2\sqrt{2}}\right){\rm cos}(at_{0}),\\
M_{3}(t_{0})&=&\int^{\infty}_{-\infty}dt\ p_{s}(t+t_{0})\ {\rm
cos}(at)=-\frac{\pi a}{\sqrt{2}}\ {\rm sech}\!\left(\frac{\pi
a}{2\sqrt{2}}\right){\rm sin}(at_{0}),
\end{eqnarray}
\end{widetext}
where we have used the fact that $p_{s}(t)$ is odd in $t$. The
Melnikov function $M(t_{0})$ for general $\vec{W}(x)$ potentials
with the small $x$ behavior as shown in Eq.~(\ref{smallx}) is
therefore an oscillatory function with infinity number of simple
zeros at large $t_{0}$. This is sufficient to indicate the
occurrence of chaotic behaviors for these general spinning
particle models.

In summary, we have indicated, using the method of Melnikov
functions, the presence of homoclinic entanglements and thus the
onset of chaos in one-dimensional spinning particle models with a
Duffing potential and some general potentials for the spins. We
expect similar results for other potentials with homoclinic or
heteroclinic orbits. Moreover, the models we consider in this
letter are derived from one-dimensional pseudoclassical systems
with three Grassmann variables, so we also expect chaos to occur
for spinning particle models derived from systems with three or
more Grassmann variables where more degrees of freedom are
involved.

As we have mentioned at the beginning, chaotic behaviors do occur
for the motions of spinning particles in a Schwarzschild black
hole spacetime which can be described by a supersymmetric theory.
This shows that we should extend our investigation of chaos in
spinning particle models to higher dimensions as well as to models
in curved spacetimes. We plan to do that in our future
publications.

We would like to thank C.-C. Chen for useful discussions. This
work was supported by the National Science Council of the Republic
of China under contract number NSC89-2112-M-032-021.

\end{document}